\documentclass{amsart}%
\usepackage{amsfonts}
\usepackage{amsmath}
\usepackage{amssymb}
\usepackage{graphicx}%
\theoremstyle{plain}

\newtheorem{corollary}{Corollary}

\newtheorem{definition}{Definition}

\newtheorem{proposition}{Proposition}
\newtheorem{remark}{Remark}

\numberwithin{equation}{section}
\begin{document}
\title[A Continuous Variable Shor Algorithm]{A Continuous Variable Shor Algorithm}
\author{Samuel J. Lomonaco, Jr.}
\address{University of Maryland Baltimore County, Baltimore, MD \ 21250, and\\
\linebreak Mathematical Sciences Research Institute, Berkeley, CA}
\email{Lomonaco@UMBC.EDU}
\urladdr{http://www.csee.umbc.edu/\symbol{126}lomonaco}
\author{Louis H. Kauffman}
\address{University of Illinois at Chicago, Chicago, IL 60607-7045}
\email{Kauffman@UIC.EDU}
\urladdr{http://www.math.uic.edu/\symbol{126}kauffman}
\thanks{This effort is partially supported by the Defense Advanced Research Projects
Agency (DARPA) and Air Force Research Laboratory, Air Force Materiel Command,
USAF, under agreement number F30602-01-2-0522, the National Institute for
Standards and Technology (NIST), and by L-O-O-P Fund Grant BECA2002. \ The
U.S. Government is authorized to reproduce and distribute reprints for
Government purposes notwithstanding any copyright annotations thereon. \ The
views and conclusions contained herein are those of the authors and should not
be interpreted as necessarily representing the official policies or
endorsements, either expressed or implied, of the Defense Advanced Research
Projects Agency, the Air Force Research Laboratory, or the U.S. Government.
\ The first author gratefully acknowledges the hospitality of the Mathematical
Science Research Institute , Berkeley, California, where some of this work was
completed. Both authors\ would also like to thank Howard Brandt and Dan
Gottesman, for some helpful conversations, and the referee for some helpful suggestions.}
\date{March 15, 2004}

\begin{abstract}
In this paper, we use the methods found in \cite{Lomonaco1} to create a
continuous variable analogue of Shor's quantum factoring algorithm. \ By this
we mean a quantum hidden subgroup algorithm that finds the period $P$ of a
function
\[
\Phi:\mathbb{R}\longrightarrow\mathbb{R}%
\]
from the reals $\mathbb{R}$ to the reals $\mathbb{R}$, where $\Phi$ belongs to
a very general class of functions, called the class of admissible functions.
\ One objective in creating this continuous variable quantum algorithm was to
make the structure of Shor's factoring algorithm more mathematically
transparent, and thereby give some insight into the inner workings of Shor's
original algorithm. \ This continuous quantum algorithm also gives some
insight into the inner workings of Hallgren's Pell's equation algorithm.

Two key questions remain unanswered. \ Is this quantum algorithm more
efficient than its classical continuous variable counterpart? \ Is this
quantum algorithm or some approximation of it implementable?

\end{abstract}
\maketitle
\tableofcontents

\section{Introduction}

\bigskip

In this paper, we create a continuous variable analogue of Shor's quantum
factoring algorithm. \ This algorithm is called a continuous variable Shor
algorithm for the following reason. \ Recall that Shor's quantum factoring
algorithm \cite{Shor2}, \cite{Shor1}, \cite{Lomonaco2} reduces the task of
factoring an integer $N$ to that of finding the period $P$ of a function
\[
\Phi:\mathbb{Z}\longrightarrow\mathbb{Z}\operatorname{mod}N
\]
from the integers $\mathbb{Z}$ to the integers $\mathbb{Z}$ modulo $N$. \ So
by a continuous variable analogue to Shor's factoring algorithm, we mean a
quantum algorithm that finds the period $P$ of a function
\[
\Phi:\mathbb{R}\longrightarrow\mathbb{R}%
\]
from the reals $\mathbb{R}$ to the reals $\mathbb{R}$. \ 

\bigskip

One of the objectives in creating this continuous variable quantum algorithm
was to make the structure of Shor's factoring algorithm more mathematically
transparent, and thereby give some insight into the inner workings of his
original quantum factoring algorithm \cite{Shor1}, \cite{Shor2}. \ This
continuous quantum algorithm also gives some insight into the inner workings
of Hallgren's Pell's equation algorithm\cite{Hallgren1}.

\bigskip

Whether or not this quantum algorithm is more efficient than its classical
continuous variable counterpart remains to be determined. \ By allowing
continuous variables, the complexity class of problems can easily change. For
more insight into this issue, we refer the reader to Bartlett et al
\cite{Bartlett1}, \cite{Bartlett2}.\ \ Moreover, the implementability of this
continuous variable quantum algorithm, or an approximation there of, also
remains to be determined. \ 

\bigskip

Continuous variable algorithms for two other quantum algorithms are to be
found in the open literature. \ A continuous variable analogue of Grover's
algorithm was constructed by Pati, Braunstein, and Lloyd in \cite{Braustein1};
and a continuous variable Deutsch-Jozsa algorithm was recently created by Pati
and Braunstein in \cite{Braustein2}.

\bigskip

There is also a great deal of literature written on many other areas of
continuous variable quantum information science.  For example, work on
continuous variable teleportation can be found in \cite{Braunstein3},
\cite{Caves1}, and \cite{Furusawa1}, on continuous variable quantum secrecy
sharing in \cite{Lance1} and \cite{Tyc1}, on continuous variable entanglement
in \cite{Pfister1}, and on continuous variable quantum error correction in
\cite{Barnes1} and \cite{Gottesman1}.

\bigskip

\section{Mathematical machinery}

\bigskip

To create a continuous variable analogue of Shor's algorithm, we will need to
make use of the mathematical machinery of \textbf{generalized functions} (also
known as \textbf{distributions}) and of \textbf{rigged Hilbert spaces} (also
known as \textbf{Gel'fand triplets}.) \ For more in depth discussions of this
mathematical machinery, we refer the reader to \cite{Bohm1}, \cite{Gadella1},
\cite{Gelfand1}, \cite{Gelfand2}, \cite{Richards1}, and \cite{Schwartz1}.

\bigskip

\subsection{Generalized functions}

\bigskip

In regard to generalized functions, the reader is no doubt familiar with one
generalized function, namely, the Dirac delta function
\[
\delta\left(  x\right)
\]
on the reals $\mathbb{R}$. \ \ We will also make use of the following
generalized function
\[
\delta_{P}\left(  x\right)  =\frac{1}{\left\vert P\right\vert }%
{\displaystyle\sum\limits_{n=-\infty}^{\infty}}
\delta\left(  x-\frac{n}{P}\right)  \text{,}%
\]
which is an infinite sum of Dirac delta functions over the lattice $\left\{
\frac{n}{P}:n\in\mathbb{Z}\right\}  $, where $P$ is a nonzero real number.

\bigskip

\subsection{Rigged Hilbert spaces}

\bigskip

We will make use of the rigged Hilbert space $\mathcal{H}_{\mathbb{R}}$ with
\textbf{orthonormal basis}
\[
\left\{  \left\vert x\right\rangle :x\in\mathbb{R}\right\}  \text{, }%
\]
where by \textbf{orthonormal} we mean there is a bracket product on
$\mathcal{H}_{\mathbb{R}}$ defined by
\[
\left\langle x\mid y\right\rangle =\delta\left(  x-y\right)  \text{ .}%
\]
The elements of $\mathcal{H}_{\mathbb{R}}$ are formal integrals of the form
\[%
{\displaystyle\int_{-\infty}^{\infty}}
dx\ f\left(  x\right)  \left\vert x\right\rangle \text{ ,}%
\]
where $f:\mathbb{R}\longrightarrow\mathbb{C}$ is a function or a generalized function.

\bigskip

For $x_{0}$ a constant, we define
\[
\left\vert x_{0}\right\rangle =%
{\displaystyle\int\limits_{-\infty}^{\infty}}
dx\ \delta\left(  x-x_{0}\right)  \left\vert x\right\rangle
\]
Since the Dirac delta function is a tempered distribution \cite{Richards1}, it
follows that
\[
\left\langle y_{0}\mid x_{0}\right\rangle =\left\{
\begin{array}
[c]{ll}%
1 & \text{if }x_{0}=y_{0}\\
& \\
0 & \text{otherwise}%
\end{array}
\right.
\]

\bigskip

\section{Fourier analysis on the real line $\mathbb{R}$}

\bigskip

Let $\Phi:\mathbb{R}\longrightarrow\mathbb{R}$ be a periodic admissible
function of minimum period $P$ from the reals $\mathbb{R}$ to the reals
$\mathbb{R}$. \ 

\bigskip

\begin{remark}
We have intentionally not defined the term `admissible,' since there are many
possible definitions of this term. \ For example, one workable definition of
an admissible function is a function that is Lebesgue integrable on every
closed subinterval of the reals $\mathbb{R}$. \ 
\end{remark}

\bigskip

We seek to define the Fourier transform of $\Phi$. \ Since $\Phi$ is in
general neither $L^{1}$ nor $L^{2}$ nor of compact support, the usual
definitions of the Fourier transform will not apply. \ So we need to be a bit
creative. \ 

\bigskip

We proceed to define the Fourier transform as follows:

\bigskip

\begin{definition}
Let $\Phi:\mathbb{R}\longrightarrow\mathbb{R}$ be a periodic admissible
function of minimum period $P$ from the reals $\mathbb{R}$ to the reals
$\mathbb{R}$. \ We interpret the standard expression $%
{\displaystyle\int\limits_{-\infty}^{\infty}}
dx\ e^{-2\pi ixy}\Phi\left(  x\right)  $ for the Fourier transform
$\widehat{\Phi}:\mathbb{R}\longrightarrow\mathbb{C}$ as the generalized
function%
\[
\widehat{\Phi}\left(  y\right)  =\delta_{P}\left(  y\right)
{\displaystyle\int\limits_{0}^{P}}
dx\ e^{-2\pi ixy}\Phi\left(  x\right)
\]
where
\[
\delta_{P}\left(  y\right)  =\frac{1}{\left\vert P\right\vert }%
{\displaystyle\sum\limits_{n=-\infty}^{\infty}}
\delta\left(  y-\frac{n}{P}\right)  \text{ ,}%
\]
and where $\mathbb{C}$ denotes the complex numbers.
\end{definition}

\bigskip

\begin{remark}
The above definition can be motivated as follows:
\[%
\begin{array}
[c]{ccl}%
{\displaystyle\int\limits_{-\infty}^{\infty}}
dx\ e^{-2\pi ixy}\Phi\left(  x\right)   & = &
{\displaystyle\sum\limits_{n=-\infty}^{\infty}}
{\displaystyle\int\limits_{nP}^{\left(  n+1\right)  P}}
dx\ e^{-2\pi ixy}\Phi\left(  x\right)  \\
&  & \\
& = &
{\displaystyle\sum\limits_{n=-\infty}^{\infty}}
{\displaystyle\int\limits_{0}^{P}}
dx\ e^{-2\pi i\left(  x+nP\right)  y}\Phi\left(  x+nP\right)  \\
&  & \\
& = &
{\displaystyle\sum\limits_{n=-\infty}^{\infty}}
e^{-2\pi inPy}%
{\displaystyle\int\limits_{0}^{P}}
dx\ e^{-2\pi ixy}\Phi\left(  x\right)  \\
&  & \\
& = &
{\displaystyle\sum\limits_{n=-\infty}^{\infty}}
\frac{1}{\left\vert P\right\vert }\delta\left(  y-\frac{n}{P}\right)
{\displaystyle\int\limits_{0}^{P}}
dx\ e^{-2\pi ixy}\Phi\left(  x\right)  \\
&  & \\
& = & \delta_{P}\left(  y\right)
{\displaystyle\int\limits_{0}^{P}}
dx\ e^{-2\pi ixy}\Phi\left(  x\right)
\end{array}
\]
where, in the context of distributions, we have
\[%
{\displaystyle\sum\limits_{n=-\infty}^{\infty}}
e^{-2\pi inPy}=\frac{1}{\left\vert P\right\vert }\delta\left(  y-\frac{m}%
{P}\right)  \text{ , for }y\in\left[  \frac{m}{P},\frac{m+1}{P}\right)
\]
(See \cite{Richards1}.)
\end{remark}

\bigskip

The reader can easily verify that the inverse Fourier transform behaves as
expected, i.e., that

\bigskip

\begin{proposition}%
\[
\Phi\left(  x\right)  =%
{\displaystyle\int\limits_{-\infty}^{\infty}}
dy\ e^{-2\pi ixy}\widehat{\Phi}\left(  y\right)
\]

\end{proposition}

\bigskip

\section{The algorithm for finding integer periods}

\bigskip

Let
\[
\Phi:\mathbb{R}\longrightarrow\mathbb{R}%
\]
be a periodic admissible function of minimum period $P$ from the reals
$\mathbb{R}$ to the reals $\mathbb{R}$. \ We will now create a continuous
variable Shor algorithm to find integer periods. \ In later sections, we will
extend the algorithm to rational periods, and then to irrational periods.

\bigskip

We construct two quantum registers \bigskip%

\[
\left\vert \text{\textsc{Left Register}}\right\rangle \text{ and }\left\vert
\text{\textsc{Right Register}}\right\rangle
\]
called left- and right-registers respectively, each `living' respectively in
its own separate rigged Hilbert space $\mathcal{H}_{\mathbb{R}}$. \ The left
register is constructed to hold arguments of the function $\Phi$, the right to
hold the corresponding function values.

\bigskip

We assume we are given the unitary transformation
\[
U_{\Phi}:\mathcal{H}_{\mathbb{R}}\otimes\mathcal{H}_{\mathbb{R}}%
\longrightarrow\mathcal{H}_{\mathbb{R}}\otimes\mathcal{H}_{\mathbb{R}}%
\]
defined by
\[
U_{\Phi}:\left\vert x\right\rangle \left\vert y\right\rangle \longmapsto
\left\vert x\right\rangle \left\vert y+\Phi\left(  x\right)  \right\rangle
\]

\bigskip

Finally, we choose a large positive integer $Q$, so large that $Q\geq2P^{2}$.

\bigskip

The quantum part of our algorithm consists of \fbox{\textbf{Step 0}} through
\fbox{\textbf{Step 4}} as described below:

\bigskip

\begin{itemize}
\item[\fbox{\textbf{Step 0}}] Initialize
\[
\left\vert \psi_{0}\right\rangle =\left\vert 0\right\rangle \left\vert
0\right\rangle
\]

\item[\fbox{\textbf{Step 1}}] Apply the inverse Fourier transform to the left
register, i.e. apply $\mathcal{F}^{-1}\otimes1$ to obtain
\[
\left\vert \psi_{1}\right\rangle =%
{\displaystyle\int\limits_{-\infty}^{\infty}}
dx\ e^{2\pi ix\cdot0}\left\vert x\right\rangle \left\vert 0\right\rangle =%
{\displaystyle\int\limits_{-\infty}^{\infty}}
dx\ \left\vert x\right\rangle \left\vert 0\right\rangle
\]

\item[\fbox{\textbf{Step 2}}] Apply $U_{\Phi}:\left\vert x\right\rangle
\left\vert u\right\rangle \longmapsto\left\vert x\right\rangle \left\vert
u+\Phi\left(  x\right)  \right\rangle $ to obtain
\[
\left\vert \psi_{2}\right\rangle =%
{\displaystyle\int\limits_{-\infty}^{\infty}}
dx\ \left\vert x\right\rangle \left\vert \Phi\left(  x\right)  \right\rangle
\]

\item[\fbox{\textbf{Step 3}}] Apply the Fourier transform to the left
register, i.e. apply $\mathcal{F}\otimes1$ to obtain
\begin{align*}
\left\vert \psi_{3}\right\rangle  &  =%
{\displaystyle\int\limits_{-\infty}^{\infty}}
dy%
{\displaystyle\int\limits_{-\infty}^{\infty}}
dx\ e^{-2\pi ixy}\left\vert y\right\rangle \left\vert \Phi\left(  x\right)
\right\rangle =%
{\displaystyle\int\limits_{-\infty}^{\infty}}
dy\ \left\vert y\right\rangle \sum_{n=-\infty}^{\infty}%
{\displaystyle\int\limits_{nP}^{\left(  n+1\right)  P}}
dx\ e^{-2\pi ixy}\left\vert \Phi\left(  x\right)  \right\rangle \\
& \\
&  =%
{\displaystyle\int\limits_{-\infty}^{\infty}}
dy\ \left\vert y\right\rangle \sum_{n=-\infty}^{\infty}%
{\displaystyle\int\limits_{0}^{P}}
dx\ e^{-2\pi i\left(  x+nP\right)  y}\left\vert \Phi\left(  x+nP\right)
\right\rangle \\
& \\
&  =%
{\displaystyle\int\limits_{-\infty}^{\infty}}
dy\ \left\vert y\right\rangle \left(  \sum_{n=-\infty}^{\infty}e^{-2\pi
inPy}\right)  \left(
{\displaystyle\int\limits_{0}^{P}}
dx\ e^{-2\pi ixy}\left\vert \Phi\left(  x\right)  \right\rangle \right) \\
& \\
&  =%
{\displaystyle\int\limits_{-\infty}^{\infty}}
dy\ \left\vert y\right\rangle \delta_{P}\left(  y\right)  \left(
{\displaystyle\int\limits_{0}^{P}}
dx\ e^{-2\pi ixy}\left\vert \Phi\left(  x\right)  \right\rangle \right)  =%
{\displaystyle\sum\limits_{n=-\infty}^{\infty}}
\left\vert \frac{n}{P}\right\rangle \left(  \frac{1}{\left\vert P\right\vert }%
{\displaystyle\int\limits_{0}^{P}}
dx\ e^{-2\pi ix\frac{n}{P}}\left\vert \Phi\left(  x\right)  \right\rangle
\right) \\
& \\
&  =%
{\displaystyle\sum\limits_{n=-\infty}^{\infty}}
\left\vert \frac{n}{P}\right\rangle \left\vert \Omega\left(  \frac{n}%
{P}\right)  \right\rangle
\end{align*}

\end{itemize}

\bigskip where
\[
\left\vert \Omega\left(  \frac{n}{P}\right)  \right\rangle =\frac
{1}{\left\vert P\right\vert }%
{\displaystyle\int\limits_{0}^{P}}
dx\ e^{-2\pi ix\frac{n}{P}}\left\vert \Phi\left(  x\right)  \right\rangle
\text{ .}%
\]

\bigskip

\begin{itemize}
\item[\fbox{\textbf{Step 4}}] Measure the left register with respect to the
observable
\[
\mathcal{O}=%
{\displaystyle\int\limits_{-\infty}^{\infty}}
dy\ \frac{\left\lfloor Qy\right\rfloor }{Q}\left\vert y\right\rangle
\left\langle y\right\vert
\]
to produce a random eigenvalue
\[
\frac{m}{Q}\text{ ,}%
\]
where $\left\lfloor Qy\right\rfloor $ denotes the greatest integer $\leq Qy$,
and then determine whether $\frac{m}{Q}$ can be used to find the period $P$.
\end{itemize}

\bigskip

\section{The observable $\mathcal{O}$}

\bigskip

In this section, we now discuss the above \fbox{\textbf{Step 4}} in greater detail.

\bigskip

The spectral decomposition of the observable $\mathcal{O}$ is given by
\[
\mathcal{O}=%
{\displaystyle\int\limits_{-\infty}^{\infty}}
dx\ \frac{\left\lfloor Qx\right\rfloor }{Q}\left\vert x\right\rangle
\left\langle x\right\vert =%
{\displaystyle\sum\limits_{m=-\infty}^{\infty}}
\left(  \frac{m}{Q}\right)  P_{m}\text{ ,}%
\]
where $P_{m}$ denotes the projection operator
\[
P_{m}=%
{\displaystyle\int\limits_{\frac{m}{Q}}^{\frac{m+1}{Q}}}
dx\ \left\vert x\right\rangle \left\langle x\right\vert
\]
Measurement of the left register of
\[
\left\vert \psi_{3}\right\rangle =%
{\displaystyle\sum\limits_{n=-\infty}^{\infty}}
\left\vert \frac{n}{P}\right\rangle \left\vert \Omega\left(  \frac{n}%
{P}\right)  \right\rangle
\]
with respect to $\mathcal{O}$ will always produce an eigenvalue $\frac{m}{Q}$
for which there exists an integer $n$ such that
\[
\frac{m}{Q}\leq\frac{n}{P}<\frac{m+1}{Q}%
\]
We seek to determine the unknown rational $\frac{n}{P}$ from the known
rational eigenvalue $\frac{m}{Q}$.

\bigskip

If \ $Q\geq2P^{2}$, then the unknown rational $\frac{n}{P}$ will be a
convergent of the continued fraction expansion of the known eigenvalue
$\frac{m}{Q}$. \ Thus, the continued fraction recursion can be used to
determine the period $P$. (See \cite[Theorem 184, Section 10.15]{Hardy1}.)

\bigskip

\section{The algorithm for finding rational periods}

\bigskip\label{rationalalgorithm}

We now extend the above algorithm to one for finding rational periods
\[
P=\frac{a}{b}\text{, \ \ }\gcd\left(  a,b\right)  =1
\]
We choose an integer $Q\geq2a^{2}$. \ 

\bigskip

\begin{itemize}
\item[\fbox{\textbf{Part 1}}] \textit{Execute the above steps }%
\fbox{\textbf{Step 0}}\textit{ through }\fbox{\textbf{Step 4}}\textit{ twice
to produce two eigenvalues}
\[
\frac{m_{1}}{Q}\text{ and }\frac{m_{2}}{Q},
\]
\textit{and then goto }\fbox{\textbf{Part 2}}.
\end{itemize}

\vspace{0.3in}

Since $Q\geq2a^{2}$, the eigenvalues $\frac{m_{1}}{Q}$ and $\frac{m_{2}}{Q}%
$\ will have unique convergents respectively of the form
\[
\frac{n_{1}b}{a}\text{ and }\frac{n_{2b}}{a}%
\]
(See \cite[Theorem 184, Section 10.15]{Hardy1}.)

\bigskip

If the following \textsc{Condition A} is satisfied, then the reciprocal period
is simply given by
\[
\frac{1}{P}=\frac{\gcd\left(  n_{1}b,n_{2}b\right)  }{a}%
\]

\vspace{0.3in}

\noindent\textsc{Condition A.} \ \fbox{$\gcd\left(  n_{1},n_{2}\right)  =1$,
\ $\gcd\left(  n_{1},a\right)  =1$, \ \ $\gcd\left(  n_{2},a\right)  =1$}

\vspace{0.3in}

If we assume that \textsc{Condition A} is satisfied, then the above expression
for the reciprocal period can be computed in \fbox{\textbf{Part 2}} given below:

\vspace{0.3in}

\begin{itemize}
\item[\fbox{\textbf{Part 2}}] \textit{Execute the following:}

\begin{itemize}
\item[\fbox{\textbf{Step 5}}] \textit{Compute all the convergents }$\left\{
\frac{p_{1k}}{q_{1k}}:k=1,2,\ldots,K\right\}  $\textit{ and }$\left\{
\frac{p_{2\ell}}{q_{2\ell}}:\ell=1,2,\ldots,L\right\}  $\textit{\ of}
$\frac{m_{1}}{Q}$\textit{ and }$\frac{m_{2}}{Q}$\textit{, respectively}

\item[\fbox{\textbf{Step 6}}] \textit{Search for denominators }$q_{1k}%
$\textit{ and }$q_{2\ell}$\textit{ which are equal}

\textsc{For} $k=1,2,\ldots,K$ \textsc{do}

\ \ \ \ \ \textsc{For} $\ell=1,2,\ldots,L$ \textsc{do}

\ \ \ \ \ \ \ \ \ \ \ \textsc{If} $q_{1k}=q_{2\ell}$ \textsc{then}

\ \ \ \ \ \ \ \ \ \ \ \ \ \ \ \ \textsc{Let} $q=q_{1k}=q_{2\ell}$\textit{ and
}$\alpha=\frac{q}{\gcd\left(  p_{1k},p_{2\ell}\right)  }$

\ \ \ \ \ \ \ \ \ \ \ \ \ \ \ \ \textsc{If} $\alpha$\textit{ is a period of
}$\Phi$ \textsc{Then}

\ \ \ \ \ \ \ \ \ \ \ \ \ \ \ \ \ \ \ \ \ \textsc{Output} $\alpha$\textit{ and
}\textsc{Stop} \ \textit{\# Period found}

\ \ \ \ \ \textsc{EndFor}

\textsc{EndFor}

\textsc{goto} \fbox{\textbf{Part 1}} \textit{\# Period not found}
\end{itemize}
\end{itemize}

\vspace{0.3in}

\fbox{\textbf{Part 2}} will find and output the period $P$ provided the output
of \fbox{\textbf{Part 1}} satisfies \textsc{Condition A}. \ From the last
corollary of the Appendix, we know this will occur after \fbox{\textbf{Part
1}} is repeated an average of $O\left(  \left(  \lg\lg a\right)  ^{2}\right)
=O\left(  \left(  \lg\lg Q\right)  ^{2}\right)  $ times. \ However, since we
do not know until the completion of \fbox{\textbf{Part 2}} whether or not the
output of \fbox{\textbf{Part 1}} satisfies Condition A , both
\fbox{\textbf{Part 1}} and \fbox{\textbf{Part 2}} need to be repeated on
average at most $O\left(  \left(  \lg\lg Q\right)  ^{2}\right)  $ to finally
find the output $P$.

\bigskip

\begin{remark}
One can quadratically speedup \fbox{\textbf{Step 6}} by taking advantage of
the fact that the convergent denominators are linearly ordered.
\end{remark}

\bigskip

\section{Finding irrational periods}

\bigskip

The above algorithm can be extended to finding, to any degree of desired
precision, the period $P$ of a periodic admissible function $\Phi$ when the
period $P$ is irrational. \ But in this case, there is a severe restrictive
condition that must be imposed on the function $\Phi$. \ Namely, we need to
assume that the function $\Phi$ is continuous. \ This continuity condition is
needed for determining whether or not a rational is sufficiently close to the
unknown irrational period.

\bigskip

\section{Conclusion}

\bigskip

The continuous variable quantum algorithm constructed in this paper does give
some insight into the inner workings of Shor's original quantum factoring
algorithm. \ Moreover, it also gives some insight into the inner workings of
Hallgren's Pell's equation algorithm\cite{Hallgren1}.

\bigskip

On the other hand, the quantum algorithm constructed in this paper\ raises
many more questions than it answers. \ Is this quantum algorithm more
efficient than its classical continuous variable counterpart? \ Can this
algorithm be implemented? \ Can an approximation of this algorithm be
implemented? \ 

\bigskip

\section{Appendix. Number theoretic probabilities.}

\bigskip

In this Appendix, we derive an asymptotic lower bound $\Omega\left(  \left(
\frac{1}{\lg\lg a}\right)  ^{2}\right)  $ on the probability that the output
of \fbox{\textbf{Part 1}} of the algorithm found in Section
\ref{rationalalgorithm} of this paper will satisfy the \textsc{Condition A}
defined within that Section.

\bigskip

\noindent\textbf{Notation Convention.} \ \textit{Throughout this section, the
symbol `}$p$\textit{' will always be used to denote a prime integer.}

\bigskip

\begin{proposition}
Let $a$ be a fixed positive integer. Then for every positive integer $N\geq
a$, if an integer $n$ is randomly chosen from the set integers
\[
\left\{  k\in\mathbb{Z}:0<k\leq N\right\}
\]
according to the uniform probability distribution, then the probability
\[
Prob_{N}\left(  \overset{}{\underset{}{\gcd}}\left(  a,n\right)  =1\right)
\]
that $n$ is relatively prime to $a$ is bounded below by
\[
Prob_{N}\left(  \overset{}{\underset{}{\gcd}}\left(  a,n\right)  =1\right)
\geq\frac{\varphi\left(  a\right)  }{a}\text{, }%
\]
where $\mathbb{Z}$ denotes the set of integers, and where $\varphi$ denotes
the Euler phi function.
\end{proposition}

\begin{proof}%
\[
Prob_{N}\left(  \overset{}{\underset{}{\gcd}}\left(  a,n\right)  =1\right)  =%
{\displaystyle\prod\limits_{p\mid a}}
\left(  1-\frac{\left\lfloor N/p\right\rfloor }{N}\right)  \geq%
{\displaystyle\prod\limits_{p\mid a}}
\left(  1-\frac{1}{p}\right)  =\frac{\varphi\left(  a\right)  }{a}%
\]

\end{proof}

\bigskip

As a corollary, we have:

\bigskip

\begin{corollary}
Let $a$ be a fixed positive integer. Then for every positive integer $N\geq
a$, if $n_{1}$ and $n_{2}$ are two random integers chosen independently with
replacement from the set integers
\[
\left\{  k\in\mathbb{Z}:0<k\leq N\right\}
\]
according to the uniform probability distribution, then the probability
\[
Prob_{N}\left(  \overset{}{\underset{}{\gcd}}\left(  a,n_{1}\right)
=1=\gcd\left(  a,n_{2}\right)  \right)
\]
that both $n_{1}$ and $n_{2}$ are relatively prime to $a$ is bounded below by
\[
Prob_{N}\left(  \underset{}{\overset{}{\gcd}}\left(  a,n_{1}\right)
=1=\gcd\left(  a,n_{2}\right)  \right)  \geq\left(  \frac{\varphi\left(
a\right)  }{a}\right)  ^{2}\text{, }%
\]
where $\mathbb{Z}$ denotes the set of integers, and where $\varphi$ denotes
the Euler phi function.
\end{corollary}

\bigskip

\begin{proposition}
Let $a$ be a fixed positive integer. Then for every positive integer $N\geq
a$, if $n_{1}$ and $n_{2}$ are two random integers chosen independently with
replacement from the set of integers
\[
\left\{  k\in\mathbb{Z}:0<k\leq N\right\}
\]
according to the uniform probability distribution, then the conditional
probability
\[
Prob_{N}\left(  \overset{}{\underset{}{\gcd}}\left(  n_{1},n_{2}\right)
=1\mid\gcd\left(  a,n_{1}\right)  =1=\gcd\left(  a,n_{2}\right)  \right)
\]
that $n_{1}$ and $n_{2}$ are relatively prime given that $n_{1}$ and $n_{2}$
are both relatively prime to $a$ is bounded below by
\[
Prob_{N}\left(  \overset{}{\underset{}{\gcd}}\left(  n_{1},n_{2}\right)
=1\mid\gcd\left(  a,n_{1}\right)  =1=\gcd\left(  a,n_{2}\right)  \right)
\geq\frac{6}{\pi^{2}}\text{, }%
\]
where $\mathbb{Z}$ denotes the set of integers, and where $\varphi$ denotes
the Euler phi function.
\end{proposition}

\begin{proof}%
\begin{align*}
Prob_{N}\left(  \left.  \underset{\mathstrut}{\overset{\mathstrut}{\gcd\left(
n_{1},n_{2}\right)  =1}}\right\vert \ \
\begin{array}
[c]{c}%
\gcd\left(  a,n_{1}\right)  =1\\
\text{and}\\
\gcd\left(  a,n_{2}\right)  =1
\end{array}
\right)   &  =%
{\displaystyle\prod\limits_{\underset{p\nmid a\text{ and }p\leq N}{p}}}
\left(  1-\left(  \frac{\left\lfloor N/p\right\rfloor }{N}\right)  ^{2}\right)
\\
& \\
&  \geq%
{\displaystyle\prod\limits_{\underset{p\nmid a\text{ and }p\leq N}{p}}}
\left(  1-p^{-2}\right) \\
&  >%
{\displaystyle\prod\limits_{p}}
\left(  1-p^{-2}\right)  =\zeta\left(  2\right)  ^{-1}=\frac{6}{^{\pi^{2}}}%
\end{align*}
where $\zeta$ denotes the Riemann zeta function. (See \cite{Hardy1}.)
\end{proof}

\bigskip

\begin{corollary}
Let $a$ be a fixed positive integer. Then for every positive integer $N\geq
a$, if $n_{1}$ and $n_{2}$ are two random integers chosen independently with
replacement from the set of integers
\[
\left\{  k\in\mathbb{Z}:0<k\leq N\right\}
\]
according to the uniform probability distribution, then the probability
\[
Prob_{N}\left(  \overset{}{\underset{}{\gcd}}\left(  n_{1},n_{2}\right)
=\gcd\left(  a,n_{1}\right)  =\gcd\left(  a,n_{2}\right)  =1\right)
\]
that the integers $a$, $n_{1}$, $n_{2}$ are all relatively prime to each other
is bounded below by
\[
Prob_{N}\left(  \overset{}{\underset{}{\gcd}}\left(  n_{1},n_{2}\right)
=\gcd\left(  a,n_{1}\right)  =\gcd\left(  a,n_{2}\right)  =1\right)  \geq
\frac{6}{\pi^{2}}\left(  \frac{\varphi\left(  a\right)  }{a}\right)
^{2}\text{, }%
\]
where $\mathbb{Z}$ denotes the set of integers, and where $\varphi$ denotes
the Euler phi function. \ Moreover, we have the asymptotic bound
\[
Prob_{N}\left(  \overset{}{\underset{}{\gcd}}\left(  n_{1},n_{2}\right)
=\gcd\left(  a,n_{1}\right)  =\gcd\left(  a,n_{2}\right)  =1\right)
=\Omega\left(  \left(  \frac{1}{\lg\lg a}\right)  ^{2}\right)
\]

\end{corollary}

\begin{proof}
The first part of this corollary follows immediately from the above corollary
and proposition. \ The second part follows immediately from a number theoretic
theorem found in \cite[Theorem 328, Section 18.4]{Hardy1} which states that
\[
\lim\inf\frac{\varphi\left(  a\right)  }{a/\ln\ln a}=e^{-\gamma}\text{,}%
\]
where $\gamma$ denotes Euler's constant.
\end{proof}

\end{document}